\documentclass{article}

\usepackage{arxiv}

\usepackage[utf8]{inputenc} 
\usepackage[T1]{fontenc}    
\usepackage{hyperref}       
\usepackage{url}            
\usepackage{booktabs}       
\usepackage{amsfonts}       
\usepackage{nicefrac}       
\usepackage{microtype}      
\usepackage{lipsum}		
\usepackage{graphicx}
\usepackage{natbib}
\usepackage{doi}

\title{Enhancing Programming Education with ChatGPT: A Case Study on Student Perceptions and Interactions in a Python Course}

\author{Boxuan Ma \\
	Faculty of Arts and Science\\
	Kyushu University, Japan\\
	\texttt{boxuan@artsci.kyushu-u.ac.jp} \\
	\And
	\hspace{1mm}Chen Li \\
	Faculty of Information Science and Electrical Engineering\\
	Kyushu University, Japan\\
	\texttt{chenli@limu.ait.kyushu-u.ac.jp} \\
	\And
	\hspace{1mm}Konomi Shin'ichi \\
	Faculty of Arts and Science\\
	Kyushu University, Japan\\
	\texttt{konomi@artsci.kyushu-u.ac.jp} \\
}

\renewcommand{\shorttitle}{\textit{arXiv} Template}

\begin{document}
\maketitle

\begin{abstract}

The integration of ChatGPT as a supportive tool in education, notably in programming courses, addresses the unique challenges of programming education by providing assistance with debugging, code generation, and explanations. Despite existing research validating ChatGPT's effectiveness, its application in university-level programming education and a detailed understanding of student interactions and perspectives remain limited. This paper explores ChatGPT's impact on learning in a Python programming course tailored for first-year students over eight weeks. By analyzing responses from surveys, open-ended questions, and student-ChatGPT dialog data, we aim to provide a comprehensive view of ChatGPT's utility and identify both its advantages and limitations as perceived by students. Our study uncovers a generally positive reception toward ChatGPT and offers insights into its role in enhancing the programming education experience. These findings contribute to the broader discourse on AI's potential in education, suggesting paths for future research and application.

\end{abstract}


\section{Introduction}
The integration of generative artificial intelligence into the educational landscape marks a transformative shift in how teaching and learning processes are conceptualized and delivered. Emerging as a leading model of Large-scale language models (LLMs), ChatGPT is rapidly gaining recognition for its potential to significantly enrich both teaching and learning experiences by enabling users to retrieve explanations for various concepts in just a few minutes via conversation \cite{humble2023cheaters,malinka2023educational}. Recently, ChatGPT has seen increasing utilization in education to support teachers and students, including creating educational content, improving student engagement and interaction, as well as personalizing learning experiences \cite{kasneci2023chatgpt}.

As ChatGPT is not limited to natural language but can also handle programming languages, its potential impact on programming education is significant. The use of ChatGPT in program learning has also become widespread lately, as learning programming is a challenging and complex process for most people \cite{rahman2023chatgpt,rajala2023call,yilmaz2023augmented}. It can aid students through the complexities of programming languages, such as debugging with code,  and providing real-time problem-solving assistance \cite{biswas2023role}. Numerous studies have validated GPT's performance in solving programming problems, including debugging, generating code, and providing explanations \cite{pankiewicz2023large,phung2023generating,phung2023generative,sobania2023analysis,tian2304chatgpt}. 

Although these studies provide initial insights into ChatGPT's potential and challenges, they either predominantly focus on ChatGPT's capabilities or investigate from educators' perspective rather than students', which often neglects the detailed experiences of learners. To address this gap, our study aims to examine how first-year university students utilize ChatGPT in an eight-week Python programming course. Through questionnaires and open-ended questions, we investigate students' perceptions of ChatGPT's role in their learning, highlighting its strengths and areas for enhancement. By evaluating conversation interactions for various learning activities, our research offers valuable insights for effectively integrating AI into programming education. Our findings contribute to the ongoing discussion about deploying generative AI tools in educational contexts, providing recommendations for educators and curriculum developers on leveraging ChatGPT in programming courses to maximize learning outcomes while mitigating potential drawbacks.

\section{Related Work}

In recent years, the application of Artificial Intelligence in Education has gained significant traction, aiming to enhance and innovate learning and teaching methodologies through intelligent technologies. Generative AI tools like ChatGPT have demonstrated immense potential in education, offering personalized learning experiences, student support, and innovative delivery of course content \cite{anagnostopoulos2023chatgpt,kasneci2023chatgpt}. These tools provide real-time interactive feedback based on students' learning progress and needs, greatly improving learning efficiency and engagement \cite{humble2023cheaters,luckin2016intelligence}.

In programming education, ChatGPT has shown the potential to profoundly impact students' learning experiences. As an AI language model that can interact using natural language, ChatGPT enables even those without prior programming knowledge to solve coding problems with ease, significantly lowering the barrier to learning programming \cite{surameery2023use,yilmaz2023augmented}. Recent studies have highlighted ChatGPT's powerful capabilities in various programming tasks. For instance, Tian et al. \cite{tian2304chatgpt} empirically analyzed ChatGPT's potential as a fully automated programming assistant for code generation, program repair, and code summarization. Their results demonstrated that ChatGPT effectively handles typical programming challenges, such as fixing bugs, providing descriptions, and generating code. Moreover, research has found that ChatGPT's bug-fixing performance is competitive with common deep-learning approaches and notably better than standard program repair techniques \cite{sobania2023analysis}. Phung et al. \cite{phung2023generative} systematically compared GPT models with human tutors using different Python programming problems and real-world buggy programs, revealing that GPT-4 comes close to matching human tutors' performance in several scenarios. Furthermore, ChatGPT assists in providing feedback on programming assignments, supporting students' practical application of theoretical knowledge. Pankiewicz and Baker \cite{pankiewicz2023large} employed the GPT-3.5 model to address the challenge of generating personalized feedback for programming learning. Their experiments showed that students positively rated the usefulness of GPT-generated hints. Chen et al. \cite{chen2023gptutor} proposed GPTutor, a Visual Studio Code extension using the ChatGPT API to provide programming code explanations, and received positive feedback from students and teachers.

While ChatGPT's application in programming education has received positive attention, existing research primarily focuses on assessing its performance in solving programming tasks, with less emphasis on students' actual experiences and perceptions. Although studies suggest interactions with ChatGPT afford personalized learning experiences, enhancing motivation and fostering critical thinking and problem-solving skills \cite{shoufan2023exploring,skjuve2023user,tlili2023if}, these are not specific to programming education. A few studies have explored students' programming learning experiences with ChatGPT, but either not within a course \cite{biswas2023role} or without analyzing specific interaction data between students and ChatGPT \cite{humble2023cheaters,yilmaz2023augmented}. To address this gap, our study aims to deeply explore student interactions with ChatGPT within a complete programming course, examining how these interactions affect students' learning experiences and outcomes. By collecting and analyzing dialogue data between students and ChatGPT, along with student feedback, we comprehensively assess the value of ChatGPT's application in programming education.

\begin{figure*}[tb]
\centering
\includegraphics[scale=0.6]{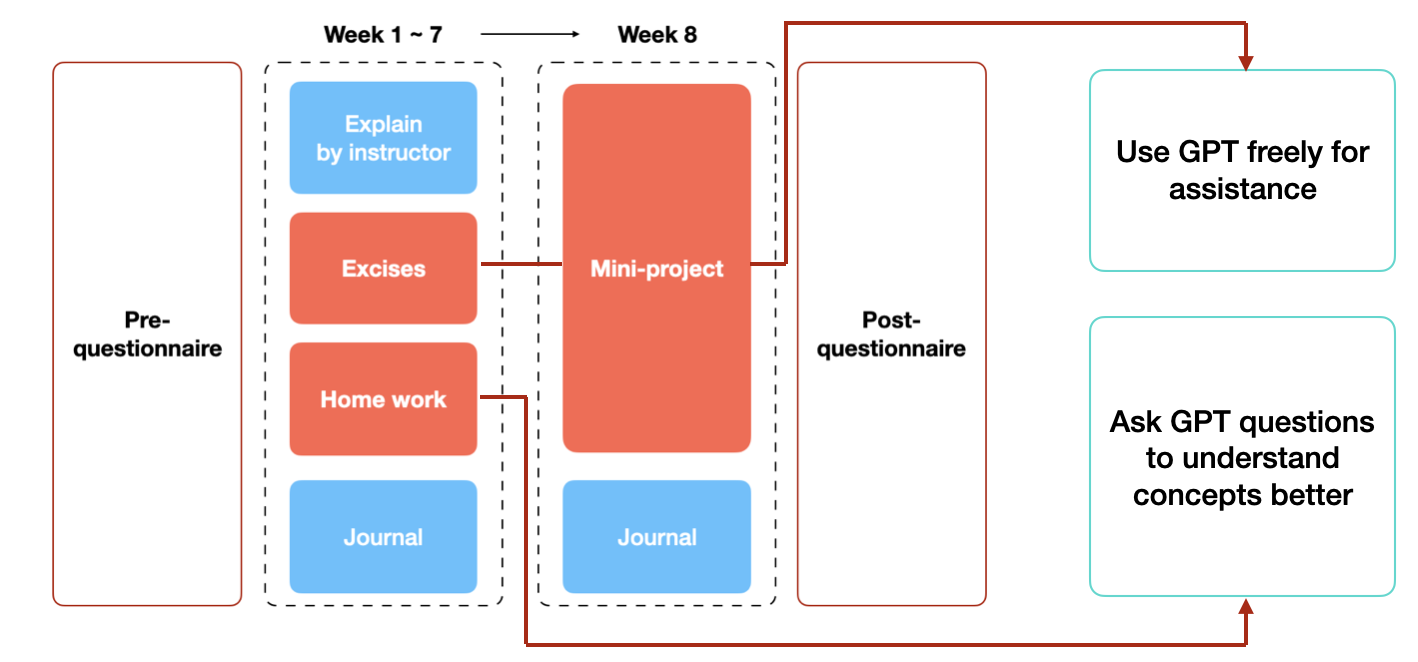}
\caption{Workflow of the study.}
\end{figure*}

\section{Method}

\begin{table*}[tb]
\setlength{\tabcolsep}{14.5mm}{
\centering
\renewcommand{\arraystretch}{1.2}  
\caption{Pre-questionnaire items (Originally in Japanese, we translate it into English).}
\begin{tabular}{cl} \hline
No. & Item \\ \hline
Q1& How familiar are you with ChatGPT?\\
Q2 &Do you use ChatGPT?\\
Q3 &What is your level in Python programming?\\
Q4 &Do you think using ChatGPT to learn Python programming is beneficial \\
& or detrimental?\\
Q5 & In your own words, what is ChatGPT? (Open-ended Question)\\ \hline
\end{tabular}}
\end{table*}

This study aims to investigate students' experiences with ChatGPT within a university Python programming exercise course. The study employs both quantitative and qualitative methods. The data collection process involved pre-questionnaire and post-questionnaire, including open-ended questions designed by the researchers. We set up the questionnaire on Google Forms and shared the online form link with the students, asking them to complete it. Students are also allowed to use ChatGPT freely to complete different activities throughout the class. They can log into a web-based interface that leverages the GPT-4 API, allowing them to engage in conversations directly through this platform. All the conversations between students and ChatGPT were collected for further analysis.

\subsection{Participants and Context}
The study was conducted among 26 students (69\% males) who were enrolled in a Python programming exercises course at our university during Winter 2023, and all participants were undergraduate students from Japan. IRB approval was secured for the study.  

The course teaches the foundations of Python programming language over an eight-week period. Each week, students attend two consecutive lessons, each lesson lasting 90 minutes and structured with the first 45 minutes dedicated to teaching (presentations and e-books were also utilized), followed by 45 minutes for students to work on programming assignments related to that week's topic. This setup aimed to enable students to practice the theoretical concepts they learned. Students can submit their tasks through the Learning Management System (LMS), which automatically logs and checks their submissions and keeps a record of all interactions and each code submission.

\subsection{Process and Data Collection}

Figure 1 shows an overview of this study. The course has three different learning activities focused on programming exercises, concept mastery, and problem-solving. Students are allowed to use ChatGPT freely to complete these activities. Pre- and post-questionnaires were also conducted to understand students' experiences with ChatGPT better. 

\textbf{Pre-questionnaire} At the beginning of the course, we asked all participants to fill out a pre-questionnaire to collect demographics such as major and gender. As shown in Table 1, we also asked about their programming experiences and their familiarity with ChatGPT. Moreover, we asked the participants to describe ChatGPT in their own words using an open-ended question.

\textbf{Programming Assignment:} Students need to complete programming assignments related to that week's topic (e.g., write a specific function based on the requirements) and submit their solutions to the LMS to verify correctness. For this activity, students are allowed to use ChatGPT freely, from obtaining code explanations and code examples to seeking debugging assistance and optimization tips. After this, students were instructed to record their dialogue data with GPT and submit them to LMS.

\begin{table*}[tb]
\centering
\caption{Post-questionnaire items (Originally in Japanese, we translate it into English).}
{\small 
\setlength{\tabcolsep}{4.5mm} 
\renewcommand{\arraystretch}{1} 
\begin{tabular}{cl} \hline
No.& Item \\ \hline
Q1 & Is ChatGPT helpful for learning programming?\\
Q2 & Will you keep using ChatGPT for learning programming?\\
Q3 & Will you use ChatGPT often?\\
Q4 & Will you recommend ChatGPT to friends?\\
Q5 & How do you think ChatGPT can help with programming learning? (Multiple answers allowed)\\
Q6 & In your own words, what is ChatGPT after actually using it? (Open-ended Question)\\
Q7 & What are the advantages of using ChatGPT for programming learning? (Open-ended Question)\\
Q8 & What are the limitations or disadvantages of using ChatGPT for programming learning? (Open-ended Question)\\
Q9 & How could ChatGPT be improved to better assist with programming learning? (Open-ended Question)\\
\hline
\end{tabular}
} 
\end{table*}

\textbf{Homework}:  At the end of each class, students were assigned homework that asked them to use ChatGPT to investigate topics that were not fully understood during class or explore subjects of personal interest not covered in the lectures. By interacting with ChatGPT, students are expected to gain deeper insights into complex subjects, receive tailored explanations, and expand their knowledge base in areas that intrigue them. In addition, students needed to provide summaries of the topics they explored using their own words. 

\textbf{Individual Mini-projects:} In the final week, students were asked to complete an individual project. Students were encouraged to apply the knowledge gained throughout the course to create a unique program, which typically involved writing less than 1,000 lines of code, with the option to explore any theme, ranging from games to chatbots. They were instructed to incorporate various programming concepts covered in the course, including different data objects, conditional statements (if-else), and loops (for, while) to show their comprehensive understanding and creativity. Students can use ChatGPT freely, but the dialog data must be uploaded with their code and report. In activities like these, ChatGPT can be utilized not only for debugging, explaining, and optimization but also as a tool for brainstorming ideas, clarifying project requirements, and deepening understanding of the technical aspects of the project.

\textbf{Post-questionnaire} As the course progressed, the students' perceptions of using ChatGPT in programming education became more evident. To capture these insights, we conducted a questionnaire in the final week, gathering students' opinions to understand the advantages and challenges of integrating ChatGPT into the programming curriculum from their perspective. A 5-point Likert scale (1-completely disagree, 5-completely agree) is used to measure different aspects of the students' opinions toward using ChatGPT for learning programming. Also,  open questions were developed to determine students’ viewpoints on the use of ChatGPT for programming learning purposes.  All question items are shown in Table 2.

\section{Results}
We show our results in the following subsections. First, we present the questionnaire results. Next, we present qualitative results based on open-ended questions that provide details of participants’ interactions with ChatGPT. 

\subsection{Analysis of questionnaires}

\subsubsection{Pre-questionnaire}

The pre-questionnaire reveals that all respondents are in their first academic year, with a majority being male (68.4\%). All respondents indicated they are beginners in Python programming (Q3). In terms of familiarity with ChatGPT (Q1), a significant portion (68.4\%) reported having some knowledge, while 25\% reported low or no understanding. However, only 36.8\% of respondents are actively using ChatGPT (Q2). Finally, the perceived impact of ChatGPT on learning (Q4) was generally positive, with 78.9\% viewing it as beneficial to some extent.

 \subsubsection{Post-questionnaire}

Figure 2 presents an overview of student responses to the post-questionnaire. The results show overwhelming support for ChatGPT as a beneficial tool in learning programming, with all respondents in agreement (Q1, 75\% strongly agreeing and 25\% agreeing). Similarly, there is a high intention to continue using ChatGPT for learning programming (Q2), with 75\% strongly agreeing and 17\% agreeing to continue its use. Furthermore, 67\% of respondents plan to use ChatGPT frequently in the future (Q3, 42\% agree, and 25\% strongly agree), and the majority of respondents agree to recommend ChatGPT to their friends (Q4, 42\% strongly agreeing and 25\% agreeing). This suggests that users find value in ChatGPT's capabilities for their learning needs. Overall, the data reflects a highly positive reception of ChatGPT among learners in programming courses. 

Additionally, Figure 3 illustrates the perceived usage of ChatGPT in programming learning. The highest percentage, 26\%, indicates that ChatGPT is helpful for correcting programming code. Equal portions of respondents, at 23\% each, believe that ChatGPT aids in answering programming questions and providing code examples, highlighting its role in offering practical coding assistance and clarification on tasks. Analysis of the dialogue data shows that students frequently use ChatGPT as a debugging tool, especially when they are working on programming exercise activities. Explanation of programming concepts is also a noted benefit, with 16\% of the respondents finding it useful. Lastly, 12\% of respondents value ChatGPT for offering learning advice and resources.

\begin{figure*}[tb]
\centering
\includegraphics[scale=0.7]{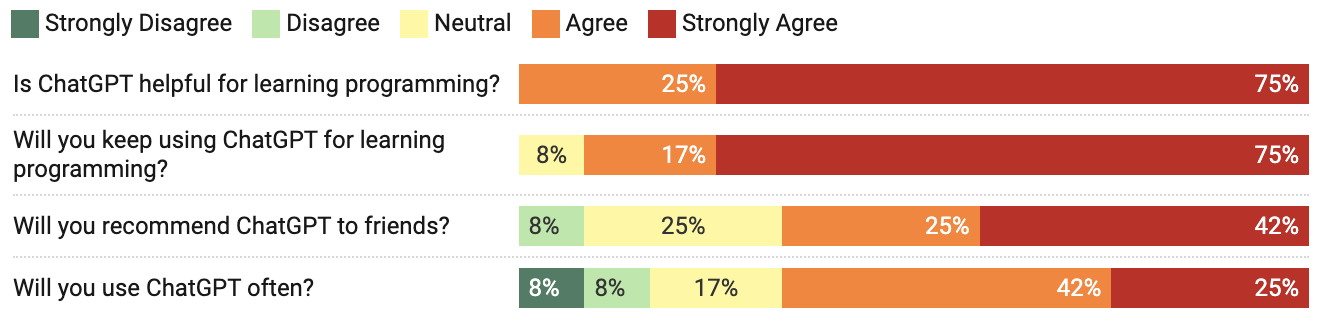}
\caption{Students’ responses to post-questionnaire (Originally in Japanese, we translate it into English.)}
\end{figure*}

\begin{figure}[tb]
\centering
\includegraphics[scale=0.5]{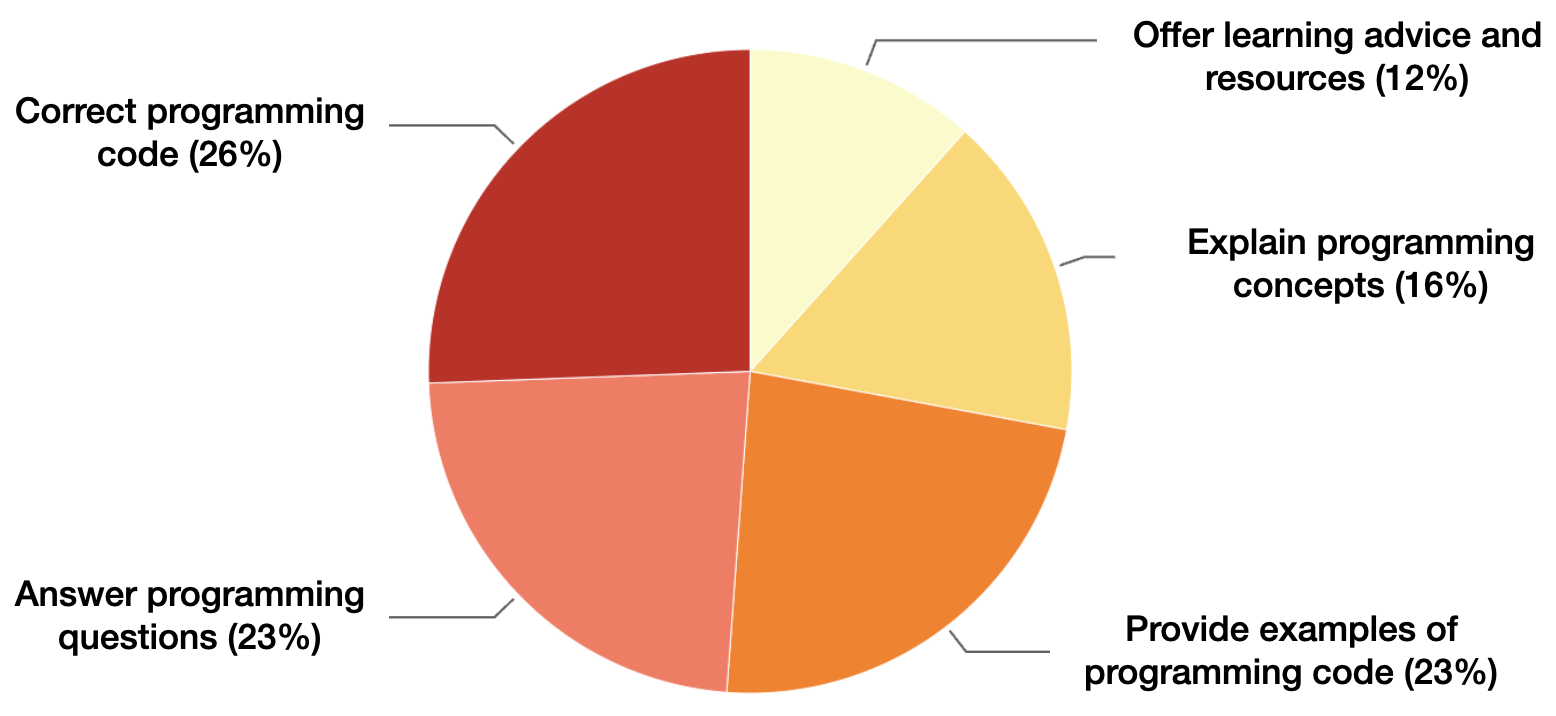}
\caption{Students' views on how ChatGPT can help with programming learning (Originally in Japanese, we translate it into English.)}
\end{figure}

\subsection{Qualitative analysis}

\begin{figure*}[tb]
\centering
\includegraphics[scale=0.55]{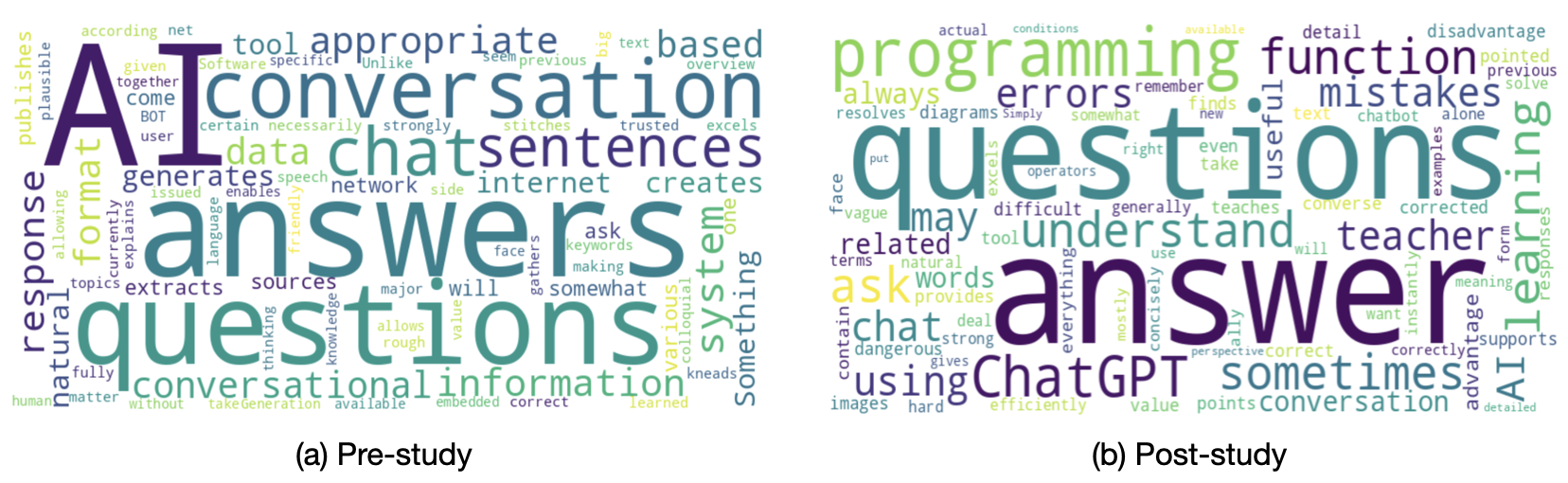}
\caption{Students’ perceptions of ChatGPT. (a) Pre-study questionnaire. (b) Post-study questionnaire. (Originally in Japanese, we translate it into English.)}
\end{figure*}

 \subsubsection{Students’ awareness on ChatGPT}

In our pre-study questionnaire (Q5 in Table 1), we asked students to describe ChatGPT in their own words. Following the learning period with ChatGPT, we repeated the question (Q6 in Table 2) in the post-study questionnaire. We then created a word cloud to visualize the shift in the students' perceptions, highlighting the most commonly used terms in their descriptions. Notably, all responses were originally in Japanese and have been translated into English for this analysis.

Figure 4 (a) shows the most frequent terms of the responses in the pre-study questionnaire, such as ``AI'', ``conversation'', ``questions'', and ``answers''. The word cloud effectively captures the essence of ChatGPT as perceived by the respondents: an AI tool designed for interactive, informative, and conversational purposes. In contrast, Figure 4 (b) from the post-study questionnaire reveals that while ``questions'' and ``answers'' remain prominent, terms like ``programming'', ``teacher'', ``learning'', and ``understand'' now figure more predominantly. 

This result reveals a significant shift in students' perceptions of ChatGPT after using it for learning purposes. Students have come to value ChatGPT more for its potential as a learning tool rather than its general conversational abilities. Moreover, students may view ChatGPT as a virtual teaching assistant, reflecting a potential change in learning habits and methodologies.

 \subsubsection{Benefits of Using ChatGPT in Programming Learning}

Students were asked to answer open questions based on their learning experience, and they highlighted several significant advantages of utilizing ChatGPT in programming education (Q7). 

It is noted that the most essential benefit of ChatGPT in this process is that it responds quickly and effectively to questions and reduces time lost in researching solutions to problems. S3 indicated, “It answers my questions about things I don’t understand, allowing me to quickly resolve my doubts.” S8 mentioned, “My questions are quickly resolved.” 

Also, many respondents like using ChatGPT to identify errors, guide corrections, and suggest improvements. Six participants explicitly mention the ability of ChatGPT to assist debugging. S1 said, “I can understand where to make corrections in the code”, and S6 indicated, “It is said that most of the time spent on programming is for fixing errors, and noticing one’s own mistakes is not easy. Asking ChatGPT to correct errors in my program allows me to rewrite it much faster than if I were to do it by myself.”

Another aspect often mentioned by the respondents is ChatGPT's ability to provide code examples and explanations, which directly helps learners apply programming knowledge to practical applications. S9's experience, where “actual examples are provided, making it very understandable and useful for application,” exemplifies how ChatGPT bridges the gap between learning and doing. S12 indicated, “I thought it would just show me examples of how to do things, but it actually points out where I’m going wrong in my programming and suggests alternative methods when I'm not satisfied with an answer, making it easy to understand.”

 \subsubsection{Limitations of Using ChatGPT in Programming Learning}
 
As for the disadvantages and limitations of using ChatGPT in the programming learning process (Q8), one of the students' points concerns the fact that using ChatGPT makes the students dependent on it, thus reducing self-thinking and learning. S8 indicated, “Because ChatGPT can give plausible answers quickly, getting into the habit of asking it anything can lead to abandoning thinking for yourself and significantly reduce learning effects.” S12 mentioned, “You lose the opportunity to think for yourself.”

Some students stated that ChatGPT may not always give the correct answers or answers they needed, particularly in the context of programming, where multiple solutions exist. S3 indicated, “You might believe incorrect information, and it's hard to notice mistakes.” Also, given that the programming course primarily covers fundamental concepts and involves relatively simple exercises, the solutions provided by ChatGPT often employ advanced and more refined methods, which can fall outside the scope of the course. S10 pointed out, “There isn't just one way to program, so ChatGPT's answers might not always be the best solution.”

 \subsubsection{Suggestions for Improving ChatGPT's Support in Programming Learning}

The feedback from users on enhancing ChatGPT for programming education is various (Q9). First, most students have highlighted the significant demand for gradual guidance, such as step-by-step hints, interactive questioning, and student-customized logical paths, to enhance the learning experience with ChatGPT. S3 suggested that “Rather than providing complete answers, giving hints step by step to encourage self-thinking would be more helpful.” S8 proposed: “Instead of just answering questions, posing appropriate questions back to the user to ensure understanding would be beneficial. This encourages users not only to receive information but also to engage in their own thought process.”

Furthermore, the need for ChatGPT to be seamlessly integrated with development environments and to offer greater accessibility was clearly stated. S7 mentioned, “Integrate with programming applications like JupyterLab to clearly display errors and solutions. Additionally, allow ChatGPT to access files on the computer when composing programs with files.” 

In addition, respondents expressed the need to ensure code correctness and the provision of diverse solutions. Feedback such as “ensuring the code provided as corrected is indeed correctly amended” (S1) and “offering multiple solutions instead of just one when presenting programming examples” (S9) was suggested to expand learners' understanding and adaptability.

\begin{figure}[tb]
\centering
\includegraphics[scale=0.2]{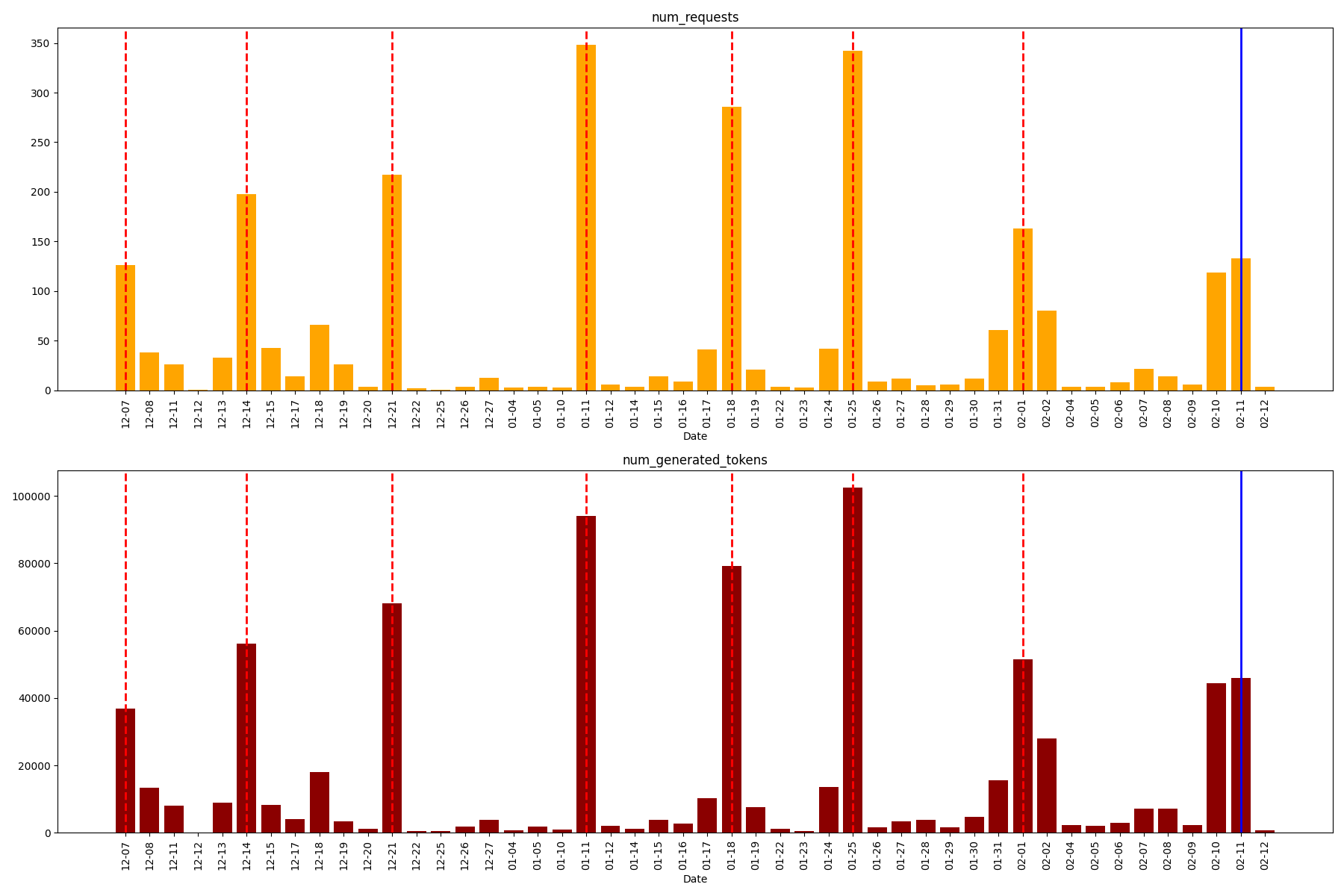}
\caption{Student-GPT interaction activity statistics.}
\end{figure}

\section{Analysis of Interactions}

Figure 5 shows the number of requests students have proposed and also the number of tokens generated by GPT models. The red dashed lines mark the days when classes were held, and the blue dashed line marks the final mini-project submission deadline. The data reflects fluctuating interaction patterns between students and GPT, which are closely aligned with the course curriculum, providing insight into the varying degrees of challenge presented by different topics. For instance, the initial few weeks showed relatively low activity, as only basic concepts were covered. However, the peaks marked periods such as weeks 4-6, where students actively sought GPT's guidance. The spike in activity during week 4 coincides with the introduction of conditional statements—a concept that likely necessitated a higher level of assistance from GPT, as evidenced by the increased number of requests and tokens. This trend was similar in weeks 5 and 6, which focused on 'for loops' and 'while loops' respectively. The complexity of these looping constructs appeared to drive a spike in GPT usage, suggesting that students were seeking more extensive support to master these more complex concepts.

\begin{figure}[tb]
\centering
\includegraphics[scale=0.25]{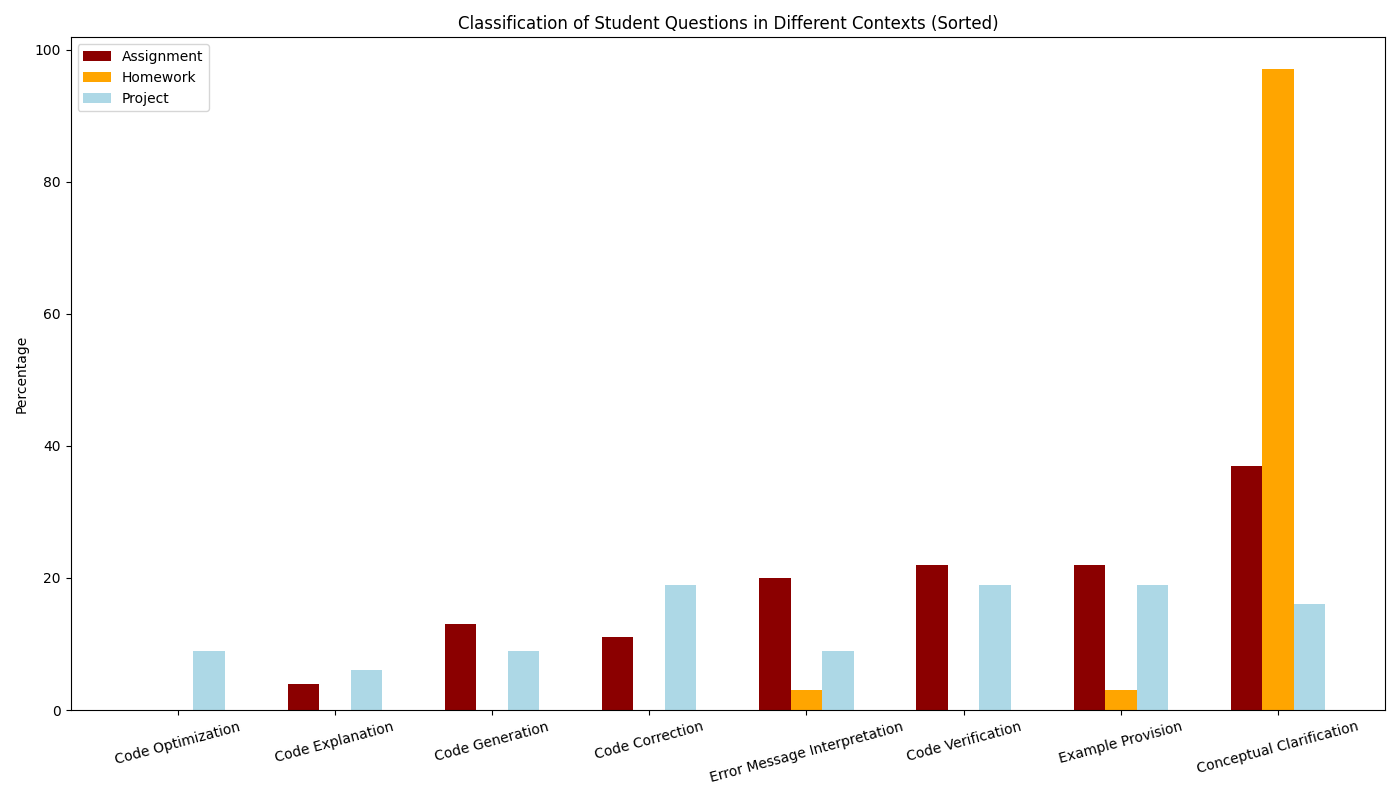}
\caption{Comparison of student question categories.}
\end{figure}

To better understand the utilization of ChatGPT across various activities, we analyzed the dialogue data between students and ChatGPT, and categorized the questions posed by students into different types. These include code verification, which involves verifying the correctness of code; Code Explanation, which entails explaining the functionality, purpose of the code; Error Message Interpretation, focused on interpreting the meaning and causes of error messages encountered during execution; Code Optimization, aimed at improving existing code; Code Correction, dedicated to debug the code; Concept Clarification, revolving around inquiring about or explaining programming-related conceptual knowledge; Code Generation, involving generating new code snippets or to fulfill specific requirements; and Example Provision, which provides example code to illustrate specific concepts or functionalities. Figure 6 shows the result of student question types. We can find that the type of questions students ask differs across the three activities. However, inquiries about conceptual understanding constitute a significant proportion of all these activities, suggesting that students are utilizing ChatGPT as a virtual teaching assistant, posing questions about concepts they do not understand and receiving timely responses. Furthermore, providing examples and code checking also represent a significant proportion.

\section{Conclusion}

This study delves into ChatGPT’s role in Python programming learning, focusing on student perceptions and interactions. Through questionnaires and open-ended questions, we explored students' perspectives on the role ChatGPT played in their learning. By evaluating students' interactions with ChatGPT, our research provides insights into effectively integrating artificial intelligence into programming education. Future work will focus on a detailed examination of students' questions and ChatGPT's answers. Evaluating student academic performance will also deepen our understanding of ChatGPT’s educational potential. Also, comparing students’ views and teachers’ views would be interesting.

\bibliographystyle{unsrtnat}
\bibliography{references}

\end{document}